\documentclass{emulateapj}
                                                                                                              
\begin{document}
\title{Altitude Limits for Rotating Vector Model Fitting of Pulsar Polarization}
\author{H. A. Craig \& Roger W. Romani}
\affiliation{Stanford University}

\begin{abstract}

Traditional pulsar polarization sweep analysis starts from the point dipole rotating vector model
(RVM) approximation. If augmented by a measurement of the sweep phase shift, one obtains
an estimate of the emission altitude (Blaskiewicz, Cordes, \& Wasserman). However, a more
realistic treatment of field line sweepback and finite altitude effects shows that this
estimate breaks down at modest altitude $\sim 0.1R_{LC}$. Such radio emission
altitudes turn out to be relevant to the young energetic and millisecond pulsars that dominate
the $\gamma$-ray population.  We quantify the breakdown height as a function of
viewing geometry and provide simple fitting formulae that allow observers to correct RVM-based
height estimates, preserving reasonable accuracy to $R\sim 0.3R_{LC}$. We discuss briefly
other observables that can check and improve height estimates.

\end{abstract}


\maketitle

\section{Introduction}

	After nearly a half century of pulsar observations, we still do not know
the detailed location of the emission zones in the neutron star magnetosphere.
However the general consensus is that the radio emission arises from the `open'
field line zone above the magnetic poles at modest altitudes, from a few to a few 
tens of neutron star radii. In contrast, the $\gamma$-ray
emission, as measured by {\it Fermi} \citep{psrcat}, is dominated by
high altitudes $> 0.1 R_{LC}$, where the light cylinder radius is $R_{LC}=cP/2\pi$.
Thus the emission zones and light curves for these two bands generally differ.
However, recently {\it Fermi} has detected $\gamma$-ray emission from a number
of millisecond pulsars where the entire magnetosphere is outside of
$R_{NS} \approx 0.2/P_{ms} R_{LC}$, so that radio emission must be from `high altitude' \citep{k12b}.
Further, \cite{kj07} and \cite{jw06} have found evidence 
that for young energetic pulsars, the radio emission is dominated by an altitude of
$\sim 1000$ km ($\sim 100 R_{NS}$). This is $\sim 0.2 R_{LC}$ for P=100 ms, and
it is precisely such young, energetic pulsars which are $\gamma$-bright.
Thus, if one is interested in $\gamma$-emitting pulsars, one must also
consider radio emission from an appreciable fraction of the light cylinder radius.

	Since the first radio observations, the high linear polarization and rapid
position angle sweep of many pulsars at cm wavelength have been used as a clue to
the geometry of the emission zone. The foundation for such study is the 
\cite{rc69} `rotating vector model' (RVM), which follows the sweep
of the magnetic field line tangent of a point dipole as projected on the sky.
Of course, finite altitude radio emission violates the point source RVM assumption and
\cite{bcw91} (hereafter BCW) gave simple approximations for the effects of
relativistic aberration at small altitude. In this approximation, the polarization
position angle is
\begin{equation}\label{eq:BCW}
\psi=\arctan\left[\frac{3 r \sin(\zeta) - \sin(\alpha) \sin(\phi+r)}
{\sin(\zeta) \cos(\alpha) -\cos(\zeta) \sin(\alpha) \cos(\phi+r)}\right],
\end{equation}
where the inclination angle between rotation axis and magnetic axis is $\alpha$, the viewing angle is 
$\zeta$, and the pulse phase is $\phi$.
The RVM formula is recovered in the limit as the scaled emission height, 
$r\equiv r_{em}/R_{LC}$, goes to zero.  
Here the principal effect is
a lag in the phase of the maximum rate of the polarization sweep ${\rm d}\psi/{\rm d}\phi|_{max}$
of $\Delta\phi \approx 2r$ from the phase of the magnetic axis. 

	If the absolute position angle of the magnetic axis on the plane of the
sky is known (eg. from the position angle of the spin axis), \cite{ha01}
show that the observed polarization gives a second height estimate,
$\Delta\psi \approx \frac{10}{3} r\cos(\alpha)$, where
\begin{equation}\label{eq:BCWPhi}
\psi=\arctan\left[\frac{-\sin(\alpha)\sin(\phi-2r)}
{\sin(\zeta) \cos(\alpha) -\cos(\zeta) \sin(\alpha) \cos(\phi-2r)}\right]+\Delta\psi
\end{equation}
\citep{d08}. In practice it is generally
unclear how to measure the magnetic axis polarization angles; most authors treat $\Delta\psi$ 
as a nuisance parameter.

	Of course, both formulae presume knowledge of the phase of closest approach 
of the magnetic axis $\phi=0$. The phase of the radio pulse peak is often used, 
but these pulses can have complex, multi-component morphology.  Further, the 
special relativistic effects shift the intensity peak forward, giving a net 
observable lag of the polarization sweep from the intensity peak 
of $\Delta \phi \approx 4r$. 
The shifts have been clarified and extended to include the effects of field 
line sweepback by \cite{dh04}, and \cite{d08}.
Nevertheless, observers generally fit to the zero altitude (RVM) limits of 
the formula to constrain $\alpha$ and $\zeta$ and, when possible,
estimate the shift of ${\rm d} \psi/{\rm d}\phi|_{max}$ to constrain the altitude,
using the linear (BCW) scaling.  While this works adequately for many non-recycled
pulsars, relatively high altitude emission is inferred for young energetic objects. For
millisecond pulsars the basic RVM model often does not fit well.

	Thus, recent strong interest in $\gamma$-ray emitting pulsars draws our
attention to objects where the radio emission may extend to 0.1$R_{LC}$ or higher,
where the standard RVM treatment is suspect. We seek here to quantify this breakdown:
if one applies an RVM/BCW fit and obtains estimates of the magnetic inclination 
angle $\alpha_f$, viewing angle $\zeta_f$, and emission height $r_f$, for what ranges
of these parameters are these fits `valid', i.e. when do the fit values and uncertainty
ranges include (at some prescribed probability) the real value
$r_r$? We develop this analysis as a guide to observers wishing to
interpret pulsar polarization data and as an indication to situations where detailed
fits to numerical models (eg. Parent et al. 2011) are required. In addition,
we suggest analytic corrections to allow useful $r_f$
estimates from simple RVM fits to extend to somewhat higher altitude.

\section{Simulation Model Assumptions}                                              

	Our approach is to use a specific 3-D magnetosphere model with plausible
radio emission zones, to `fit' the resulting light curve and polarization
sweep with the point dipole RVM formula and to parametrize the errors.
For simplicity the field lines are given by the basic swept back (retarded) dipole
popular in models of high altitude $\gamma$-ray emission \citep{rw10} and 
we assume that the radiating particle bunches follow the magnetic field lines.
In the spirit of the RVM model, we make a simple geometric construction,
projecting the field line tangent at the emission point in the lab frame
onto the plane of the sky and assume that the radio emission is polarized 
parallel to (or perpendicular to) this vector.  We do not attempt here to 
superpose multiple emission heights or to compute intrinsic polarization 
fractions. Nor do we include other physical effects such as possible 
cross-field drift of the emitting charge bunches, current-induced departure 
from the vacuum structure for energetic pulsars \citep{spit06} or higher-order 
multipole/offset dipole effects that may be important in the small 
magnetospheres of millisecond pulsars \citep{hm11}.  While our simple 
construction ignores these possible effects, we do capture the dominant 
effect of dipole sweep-back and our computed polarization sweeps pass 
smoothly to the RVM model curves at low altitude; the other physical effects 
likely only dominate very close to the light cylinder.

	We assume here that the radio emission comes from a single altitude, 
within the open zone. We then must define the open zone shape and the illumination
across it. Of course, there is a formal cap shape for the vacuum retarded 
dipole solution, where the locus of field lines tangent to
the light cylinder trace to a cap on the surface with opening angle
$\theta_R (\phi_{cap})$ varying with azimuth $\phi_{cap}$ around the magnetic axis.
Alternatively, it is common to assume a simple circular cap, with
surface angle $\theta_C (\phi_{cap})=$ constant. To roughly match the open zone
beam sizes at an emission height of 0.1$R_{LC}$ we chose a surface cap angle of $\theta_C=2^\circ$
for a neutron star of $R_{NS} = 10^{-3}R_{LC}$, i.e. a $\sim 0.2$\,s pulsar.

	For simplicity and to follow the BCW picture, we illuminate the open 
zone with a simple Gaussian profile 
\begin{equation}\label{eq:Ipulse}
I \propto e^{(\theta_{cap}/\theta_0)^2},\qquad {\rm with }\qquad \theta_0=2^\circ/{\sqrt{\ln 5}}
\end{equation}
so that the intensity falls by 5$\times$ at the `edge' of the simple circular cap.
The angles are measured at the star surface, although the corresponding radio flux
may be emitted at high altitude.
We note that there is some evidence that a conal intensity distribution
with a patchy illumination may be more typical of many pulsars \citep{lm88,kj07}. 

	To generate a model polarization sweep we select a magnetic inclination, $\alpha_r$,
and emission height, $r_{r}$. We then project the swept-back field lines 
at this altitude to the plane of sky and record the results on a 2D sky map.
Horizontal cuts across this map at a given viewing angle, $\zeta_r$, give the
polarization angle sweep, $\psi(\phi)$. We assign `measurement' errors to each value inversely proportional
to the pulse flux at its phase. We assume that the observer's integration achieves
a uniform signal-to-noise at pulse maximum, so that the polarization measurement error
there is $1^\circ$. For pulsars observed far from the magnetic axis at large
$|\beta|\equiv |\zeta-\alpha|$ this implies longer integration.
As the pulse flux falls toward the edge of the open zone the polarization
angle uncertainties increase.

\subsection{Estimating $\phi=0$}

\begin{figure}[h]
\begin{center}
\includegraphics[width=.46\textwidth]{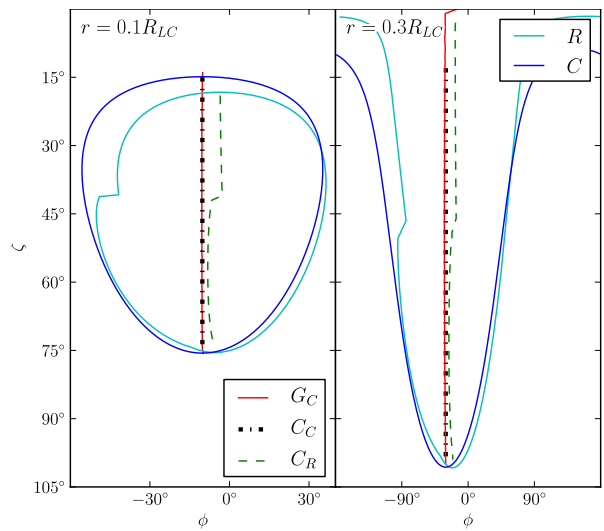}
\caption{Pulse phase estimates for $\alpha=45^\circ$, $\phi(\zeta) = 0$ 
for two different altitudes, $r=0.1R_{LC}$ and $r=0.3R_{LC}$.
$G_C$: Peak phase from maximum of simple Gaussian intensity weighting.
$C_C$: Peak phase from center of cap edges (circular cap).
$C_R$: Peak phase from center of cap edges (retarded dipole cap).}
\label{fig:Plotcap}
\end{center}
\end{figure}

	Use of the simple Gaussian illumination with the pulse phase at the
intensity peak (the projected phase of closest approach to the magnetic axis)
corresponds to the BCW assumptions. Except for very high altitude emission, where
field lines overlap in the sky map and pulse caustics can occur, this gives
a simple prescription from which $\phi=0$ may be estimated via the BCW shift. 
However, conal emission concentrated to the cap edge significantly complicates
the determination of pulse phase. One effect is the variable sweep-back at the
leading and trailing edge of the cap. Another is the particular shape of the open
zone boundary. We illustrate these effects by marking a `peak phase', the
midpoint of the projected open zone boundary, both for a simple circular cap and
for the more detailed retarded dipole cap.  Figure ~\ref{fig:Plotcap} displays 
the peak phase shifts for these different definitions.

	Not unexpectedly, Figure ~\ref{fig:Plotcap} shows that the pulse phase is 
more sensitive to the details of the open zone geometry for a conal emission zone.
The offsets shown there illustrate the effect of the retarded potential field
line flaring at high altitude. To this should be added the uncertainties associated 
with identifying the magnetic axis phase in the presence of patchy conal emission and
non-dipole field structure (for millisecond pulsars). Nevertheless, as we shall show, a substantial
fraction of the phase offset is insensitive to the choice of cap center, and can
be corrected.

\subsection{`Fitting' an RVM curve}

The retarded dipole field structure increasingly departs from the point dipole
as the scaled emission height $r=r_{em}/R_{LC}$ approaches unity. Thus
if one fits polarization data for a low altitude emitter with the RVM model, the 
fit parameters $\alpha_f,\zeta_f$, and $r_f = \Delta \phi_f/4$ at the $\chi^{2}$ minimum will be
good approximations to the real values $(\alpha_r, \zeta_r, r_r)$. For modest $r_r$
the RVM fit will absorb the sweep shape departures, (correctably) biasing the
parameter estimates, while retaining reasonable $\chi^{2}$. At large altitude, the
$\chi^{2}$ will be poor, the parameters will be uncorrectably far from the true
values and a fit to a detailed numerical model will be required. The key question
is how, with realistic errors $\sigma_{PA}$, the unabsorbed distortion grows.

	We use our estimated $\sigma_{PA}$ to construct a `$\chi^2$' weighted 
departure of the RVM model from the detailed retarded field simulation. This is the
weighted systematic error caused by the inability of the RVM model to absorb
the detailed shape of the retarded field curve. In a real observation, additional
statistical measurement errors would increase `$\chi^2$' above our model value,
especially for small $r$. Any unmodeled physical effects should additionally increase
the value of `$\chi^2$' above $\sim 1$/(degree of freedom) at the minimum.
Observers typically adopt the increase $\Delta \chi^2=\chi^2-\chi_{min}^2$ to
estimate the confidence intervals on the fit parameters. We are free to do the 
same here, since our prescription weights appropriately show where the model 
parameters are most sensitive to the data values.  We have confirmed this by 
fits to a series of Monte Carlo simulations of polarization angle data with added statistical errors,
showing that $\Delta \chi^2$ follows the usual distribution for the appropriate 
numbers of degrees of freedom. 

\section{Correcting for Bias in the RVM Height Estimates}

\begin{figure}[h]
\begin{center}
\includegraphics[width=.46\textwidth]{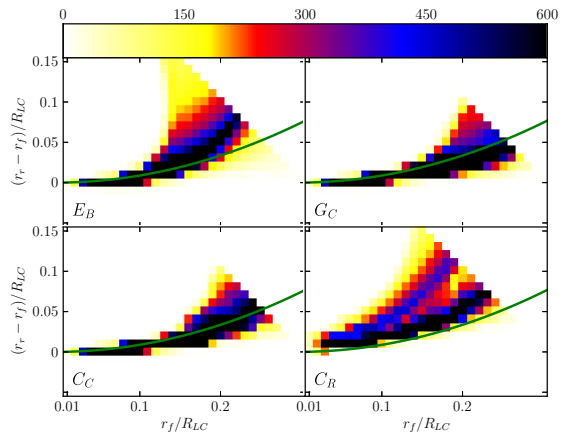}
\caption{
Altitude limits for effective RVM fits.  Each panel shows the distribution
of simulated model fits (color bar) in offset from the true altitude as a function 
of fit altitude $r_f/R_{LC}$. The dark band shows the systematic bias in the 
fit offset. The four panels are for different assumptions about the cap 
illumination and method of estimating the true phase of $\phi=0$.
$E_B$: Perfect knowledge of the location of magnetic axis in phase, without the use of an intensity model.
$G_C$: Simple Gaussian intensity peak, $\phi=0$ inferred from the altitude dependent shift of $I_{max}$. 
$C_C$: Peak intensity assigned to the center of a double pulse from edges of an open zone circular cap footpoints.
$C_R$: Peak intensity assigned to the center of a double pulse from edges of an open zone above the detailed retarded dipole cap.
The green curve shows our estimate of the bias, Equation (\ref{eq:genCor}).
}
\label{fig:totDirFitPhi}
\end{center}
\end{figure}

	Our principal goal here is to test the utility of standard RVM fits and
to provide a prescription to allow these fits wider applicability for pulsars
with high altitude emission. To do this we compare the RVM fit estimate $r_f$ with
the simulated value $r_r$. Since the mapping is not simple, statements about
ranges of validity are perforce statistical. This makes our answers mildly
sensitive to the distribution in the underlying pulsar population. Here
we assume that our parent 
pulsar population has isotropically distributed inclination and viewing angles,
ie. Prob$(\alpha)\propto {\rm sin}(\alpha)$, Prob$(\zeta) \propto {\rm sin}(\zeta)$
while the altitude is distributed uniformly on $0\le r_r\le 0.3$. Note that we
only {\it observe} a usable polarization sweep if a pulsar produces a minimum
number of phase bins (here $\Delta \phi_{obs}>0.1$). In turn,
this means that our observable pulsar population is biased toward modest
$|\beta| = |\alpha-\zeta|$.

	We generate a set of pulsar models and apply the RVM fits. This delivers
a set of observables ${\alpha_f, \zeta_f, r_f; \sigma_{\alpha_f}, \sigma_{\zeta_f}, \sigma_{r_f}}$
where the fit values are determined by $\chi^2$ minimum and the error ranges
are estimated from the curvature of the $\chi^2$ surface. An observer presented
with this set of measurements must infer the original pulsar properties.

	Focusing here on the height measurement, we test the systematic bias in
the RVM estimate. For best comparison with the BCW assumptions, we work with
the height determined from the phase lag measured from the peak of a Gaussian pulse
centered on the magnetic axis. In Figure ~\ref{fig:totDirFitPhi}, the color scale
represent the number of pulsars in the simulated population
at a given altitude derived from fitting RVM versus the difference between fitted and
real altitude. Figure ~\ref{fig:totDirFitPhi} shows that $r_f$ increasingly underestimates
$r_r$ at increasing altitude. A simple formula to provide improved height estimates 
$r_f'$ from RVM fits is then
\begin{equation}\label{eq:genCor}
r_f' = r_f+ 0.2 (r_f/0.5)^2,
\end{equation}
as plotted in Figure ~\ref{fig:totDirFitPhi}.  The line fits best to the darkest 
ridge (The ridge that contains a majority of simulated pulsars) 
for the models using the maximum of a simple Gaussian intensity peak ($G_C$)
and the center of the cap edges for a circular cap ($C_C$).  For the case using
the center of the cap edge for a retarded dipole cap ($C_R$), the line
slightly under-predicts the 
darkest ridge and does not capture the behavior of the 
second ridge which is caused by the shift of the central line from the cap notch
(see Figure ~\ref{fig:Plotcap}).  We can (unrealistically) assume that 
we know where in phase the magnetic axis 
is located and calculate the altitude from the shift in polarization directly.  Inaccuracies 
in altitude are then from Equation (\ref{eq:BCWPhi}) alone.  
With the assumption of perfect knowledge of 
the magnetic axis ($E_B$), we see the departure from the BCW formulation occurs 
at lower altitudes.  Apparently, the estimate $\Delta \phi=-2r_f$ 
for the peak intensity shift preserves good accuracy to higher altitude 
than the $\Delta \phi=+2r_f$ shift of the polarization
sweep, especially when the intensity arises from a circular cap.

\begin{figure}[h]
\begin{center}
\includegraphics[width=.4\textwidth]{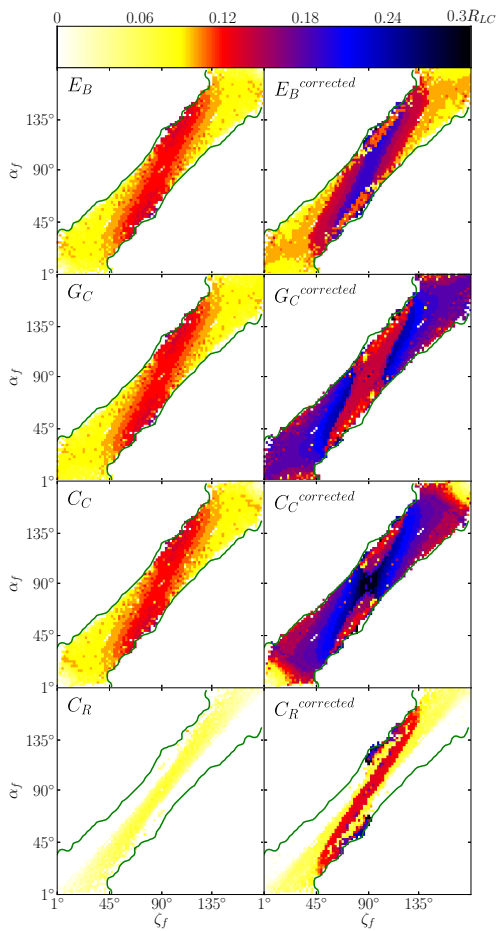}
\caption{
Maximum useful $r_f$ altitude (color bar) in the ($\alpha_f$, $\zeta_f$) plane for four
assumptions about the pulse intensity beam shape (see text for our criterion
for good fit accuracy).  Left: BCW estimates before
correction. Right: corrected heights using Equation (\ref{eq:genCor}).
Green contours indicate the area where at least fifty simulated pulsars
were fit to an ($\alpha_f$,$\zeta_f$) pair.
}
\label{fig:SmapCorrected8}
\end{center}
\end{figure}

	In practice, the height offset depends on the geometrical angles 
$(\alpha, \zeta)$. In addition, the height estimate is affected by uncertainty
in estimating the polarization sweep lag, i.e. in determining the phase of the pulse 
(or equivalently the phase of the magnetic axis). These effects are shown in 
Figure ~\ref{fig:SmapCorrected8}. For each panel we show, as a function of the estimated angles
$(\alpha_f, \zeta_f)$, the maximum height (color bar) at which the estimated altitude is
accurate. For the estimate to be useful, we require that $r_r$ lies in
the range $r_f\pm\sigma_{r_f}$ for a large fraction (99\%) of the observable
model pulsars. At small altitude this is always true. At large altitude
the distortion due to the retarded field structure causes increasing
departure from the BCW estimate. Once too small a fraction of models produce
useful fits, the BCW approximation `breaks down'.  Lowering the required fraction
does not drastically change the results seen in Figure ~\ref{fig:SmapCorrected8}, since
the fraction of failing models increases very rapidly with fit altitude. Also shown
is a green contour that marks the area where the bins contain at least
fifty simulated pulsars.  Uncolored bins are where the BCW approximation is
inaccurate at the lowest altitude.  The contours are independent 
of the intensity model (the contours are the same for each model) because
the $\alpha_f$, $\zeta_f$ bin
depends only on the polarization sweep which is calculated independently
of the intensity model.

A strong dependence between the break-down altitude and $\alpha_f$ and $\zeta_f$ exists as can 
be seen in Figure ~\ref{fig:SmapCorrected8}.  This is not due to any
difficulty in finding the phase center but arises from the nontrivial relation between
the shift in the maximum sweep of the polarization and the geometry angles.
In Figure ~\ref{fig:SmapCorrected8}, we can see that for $\alpha_f$ and $\zeta_f$
further from $90^\circ$, BCW tends to break down at a lower altitude.  
The shift in the maximum sweep of the polarization angles for these values is smaller
than predicted by the BCW model. Since the BCW model has no dependence on
$\alpha$ and $\zeta$, it is not surprising that the break-down altitude has 
a dependence on these angles.

The panels show the maximum useful height for
four different estimates of the phase lag: (top-to-bottom) perfect knowledge of
the magnetic axis, a Gaussian pulse peaked on the magnetic axis field line, a 
`conal' pulse from a field lines with a circular cap on the star and a `conal' pulse with
a cap determined by the detailed open zone of the retarded vacuum solution. 
Notice that most observed pulsars have modest $|\beta|=|\zeta-\alpha|$, and 
are close to the diagonal.
The right panels show the equivalent maximum useful height when the
estimate has been corrected according to Equation (\ref{eq:genCor}). While the uncorrected
estimates for the Gaussian pulse peak model are useful 
only to an average (over $\alpha_f$ and $\zeta_f$) height of $\overline{r_f}=0.11R_{LC}$, the corrected
estimates are usable to higher altitudes (reaching $r_f'\sim 0.3$ for the commonly
observed case of near-orthogonal rotators) with an average of $\overline{r_f}=0.22R_{LC}$.
Again, corrected
RVM estimates from a model radio pulse do better than estimates
assuming perfect knowledge of the magnetic axis, since the retarded potential
phase shifts are a fractionally larger contribution to the phase offset in this
case. 
\begin{figure}[h]
\begin{center}
\includegraphics[width=.4\textwidth]{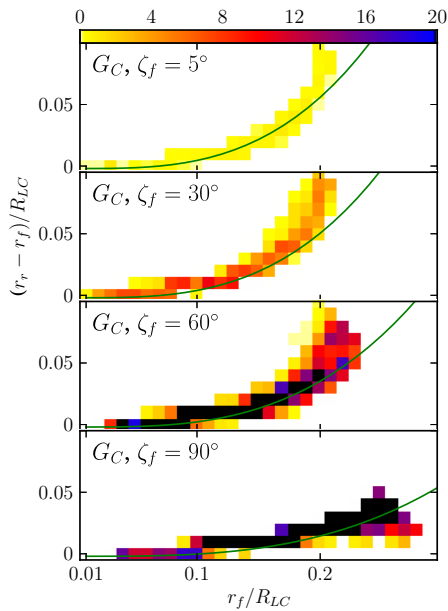}
\caption{
Altitude limits for effective RVM fits.  Each panel shows the distribution
of simulated model fits (color bar) in offset from the true altitude as a function
of fit altitude $r_f/R_{LC}$ for different $\zeta_f$, assuming a simple Gaussian intensity peak, 
$\phi=0$ inferred from the altitude dependent shift of $I_{max}$.
For these plots, the simulated pulsar population has been 
summed over $\alpha_f$ to emphasize the dominance of $\zeta_f$ in the correlation.
The green curve shows our estimate of the bias with dependence on 
$\zeta_f$, Equation (\ref{eq:zetaCor}).
}
\label{fig:zetaResmagMin}
\end{center}
\end{figure}

\begin{figure}[h]
\begin{center}
\includegraphics[width=.4\textwidth]{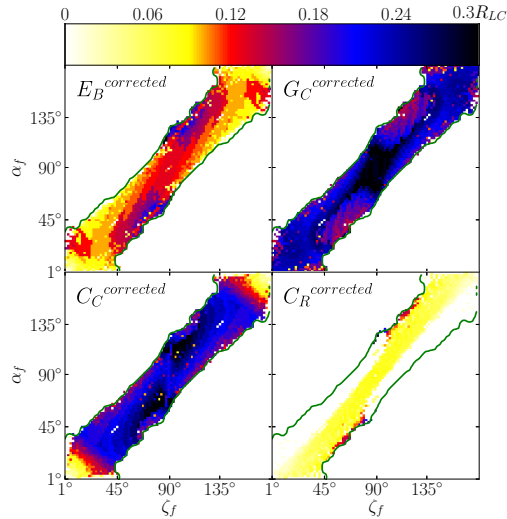}
\caption{Maximum altitude for accurate height estimates (color bar) in the 
($\alpha_f$, $\zeta_f$) plane, after applying Equation (\ref{eq:zetaCor})
(see text for our criterion for good fit accuracy).
Note that the improvement is best for a circular (Gaussian or conal) cap.
Green contours indicate the area where at least fifty simulated pulsars
were fit to an ($\alpha_f$,$\zeta_f$) pair.
}
\label{fig:SmapCorrected4}
\end{center}
\end{figure}

	We can improve the heuristic correction function by including the
viewing geometry. The bulk of the sensitivity is evidently due to $\zeta_f$,
as illustrated by the relatively small dispersion of the $r_f$ error for
individual $\zeta_f$ slices (see Figure ~\ref{fig:zetaResmagMin} for
a Gaussian central pulse).
Accordingly, we have made an alternate corrected height estimate
\begin{equation}\label{eq:zetaCor}
r_f' = r_f+ [0.3 + 0.7 |\cos(\zeta)| ] (r_f/0.5)^{3} 
\end{equation}
where $r_f=\Delta \phi/4$, as usual. This greatly extends the range for which
a simple RVM height estimate can be used (Figure ~\ref{fig:SmapCorrected4}). This estimate, based
on a Gaussian radio pulse emitted along the swept back magnetic axis, is
in general the best function for an observer to use with no other information. It provides significant
improvement in the emission height accuracy for the circular cone pulse profiles.

	Of course if one has reason to believe that a particular pulse 
profile shape is more accurate, a different correction function may be
preferred. For example if one had a double pulse arising from the open zone edges ($C_R$) and
had high confidence that this pulse filled the retarded vacuum dipole open
zone, one would correct by
\begin{equation}\label{eq:zetaCor2}
r_f' = r_f+ [0.3 +2 |\cos(\zeta)|^{2} ] (r_f/0.5)^{3}.
\end{equation}
This formulation raises the average over $\alpha_f$ and $\zeta_f$ of
the maximum useful height from $\overline{r_f}=0.05R_{LC}$ with no correction to 
$\overline{r_f}=0.15R_{LC}$.

	In general, we recommend that when an observer fits an RVM model to pulsar
data, obtaining viewing angle and polarization sweep lag measurements,
they correct their height estimate using Equation (\ref{eq:zetaCor}).  This is 
particularly useful whenever the RVM fit appears statistically adequate,
but the resulting phase lag suggests a significant emission height.
The change to the estimated height will be small for $r_f < 0.2$,
but the accuracy of the resulting estimate will be greatly increased.

	Of course, whenever $\chi^2$/DoF $\gg1$ at the fit minimum, it is a sign
that the model is inadequate. In many cases, this will be due to 
unmodeled orthogonal mode jumps and intervening scattering \citep{k09},
higher order multipoles, etc. However, for large altitudes and
multi-altitude emission the effects of sweep back and the formation
of caustics (which dominate $\gamma$-ray light curves) become dominant.
The observer should be aware that large $\chi^2$ at the fit minimum
can signal such effects and, when the inferred altitude is large, consider 
fitting the data to numerical models of 3-D pulsar magnetospheres.

\section{Height calculation from shift in $\psi$}

\begin{figure}[h]
\begin{center}
\includegraphics[width=.46\textwidth]{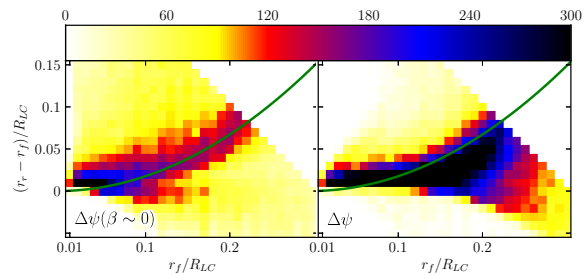}
\caption{
Altitude limits for effective RVM fits using the shift in $\psi$.  Each panel shows the distribution
of simulated model fits (color bar) in offset from the true altitude as a function
of fit altitude $r_f/R_{LC}$. The dark band shows the systematic bias in the
fit offset. 
The residual is more scattered when the altitude is measured from 
the shift in $\psi$ instead of the shift in $\phi$ of the polarization
sweep. 
On the left is the residual using the small $|\beta|$ limit.
The green curve shows our estimate of the bias from the 
shift in $\psi$, Equation (\ref{eq:DPsi1}).
}
\label{fig:totDirFitPsy}
\end{center}
\end{figure}

We can alternatively estimate $r_f$ and errors using the shift in $\psi$ \citep{ha01},
\begin{equation}\label{eq:HA}
\Delta\psi \approx \frac{10}{3} r \cos(\alpha) \left[\frac{3}{8}+\frac{5}{8} \cos(\zeta-\alpha)\right]\\
- \frac{47}{18} r \sin(\alpha) \sin(\zeta-\alpha)
\end{equation}
or, in the small $|\beta|=|\zeta-\alpha|$ limit, $\Delta\psi\approx \frac{10}{3} r \cos(\alpha)$.
As before, we compute the residual, $r_r-r_f$, as a function of $\alpha_f$, $\zeta_f$, 
and $r_f$. To estimate an emission height from the polarization shift in $\phi$,
one needs an estimate for $\phi=0$, e.g. from a pulse peak intensity model; no such intensity
model is needed if we have a measurable shift in $\psi$.
The increase with $r_f$ are shown in Figure ~\ref{fig:totDirFitPsy},
where the left panel uses the small $\beta$ limit while the right uses the full formula.
As for the $\Delta\phi$ estimate, the errors increase with $r_f$. However here, even
when the full equation (\ref{eq:HA}) is used, the corrections show a substantial spread.
In fact the uncorrected formula proves accurate ($|r_f-r_r|$ within $\sigma_{r_f}$ 99\% of the
time) only for $\overline{r_f} < 0.08$ (where $\overline{r_f}$ is again 
the average over $\alpha_f$ and $\zeta_f$)
and for $\zeta_f <60^\circ$ or $\zeta_f >120^\circ$. 
For near-orthogonal rotators the estimate is unreliable at the lowest altitudes.

	A heuristic correction to the $\Delta \psi$ estimate for Equation (\ref{eq:HA}) can be made
for $\zeta_f <60^\circ$ or $\zeta_f >120^\circ$
\begin{equation}\label{eq:DPsi1}
r_f' = r_f+ 0.4 (r_f/0.5)^{2}
\end{equation}
which allows accurate estimates to $\overline{r_f}=0.12R_{LC}$. Including the $\zeta$ dependence,
\begin{equation}\label{eq:DPsi2}
r_f' = r_f+ [0.2 +0.1 |\cos(\zeta)|^{2}] (r_f/0.5)^{2}
\end{equation}
raises the useful range to $\overline{r_f}=0.18R_{LC}$. Considering that the correction 
for the common orthogonal rotator case is especially poor, and that it is often 
difficult to infer the intrinsic $\psi_0$,
height estimates from the phase shift remain much more useful.

\section{Pulse Width Dependence on Emission Height}\label{sec:rW}

\begin{figure}[h]
\begin{center}
\includegraphics[width=.46\textwidth]{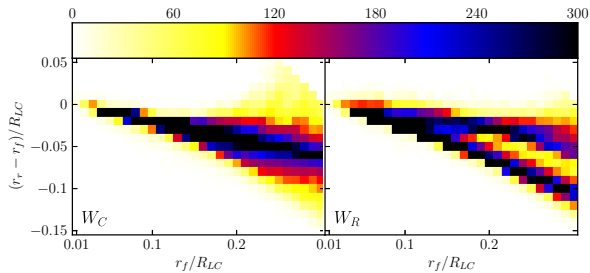}
\caption{
Altitude limits for effective pulse width.  Each panel shows the distribution
of simulated model fits (color bar) in offset from the true altitude as a function
of fit altitude $r_f/R_{LC}$. 
$W_C$: Circular cap. $W_R$: Retarded dipole cap.
The simple static dipole formula overestimates the altitude needed 
to accommodate a given pulse with $W$ in the open zone. The error depends
on the viewing geometry $\alpha$ and $\zeta$, and bifurcates for the `notched'
cap of the formal retarded potential open zone.
}
\label{fig:totDirFitW}
\end{center}
\end{figure}

	Since the field lines flare in the open zone, the full phase width $W$ of the
observed radio pulse can also be checked against the expected radio emission
altitude. The standard prescription assumes a circular cap and static dipole field
lines to infer a minimum height
\begin{equation}\label{eq:rW}
r_{W}=\frac{4}{9}\arccos^{2}\left[\cos(\alpha)\cos(\zeta)+\sin(\alpha)\sin(\zeta)\cos\left(\frac{W}{2}\right)\right].
\end{equation}
In Figure ~\ref{fig:totDirFitW} we show that the retarded dipole field flares
{\it more} than predicted by this simple formula and hence the minimum height
in Equation (\ref{eq:rW}) is an {\it over}-estimate. Thus, in general, lower
altitudes are consistent with a given observed pulse width than suggested by this
formula. Moreover, we expect that the general effect of currents in the magnetosphere
will be to increase the foot-point angles of the open zone.
This further increases the allowed
$W$ at a given height.

In general, larger widths are still most easily accommodated at large $r$ or 
small $\alpha$, but sweep-back and magnetospheric currents substantially weaken 
the minimum altitude constraints from the commonly used Equation (\ref{eq:rW}). 
Given the large sensitivity to the details of the open zone volume and the presently
unknown effect of magnetospheric currents, it is not worth developing corrections to
this formula.

\section{Conclusions; Examples from Literature}

	We conclude by examining a few RVM/BCW estimates of
emission height present in the literature.

In \cite{r11}, $\Delta \phi$ estimates were used to suggest
large emission heights for two young energetic pulsars. For J0538+2817 the
shift gives $r_f=0.15R_{LC}$, but RVM fitting only weakly constraints $\zeta$.
Applying Equation (\ref{eq:genCor}), we would infer $r^\prime_f = 0.17 R_{LC}$,
a small, but significant increase which makes it easier to accommodate the
large observed pulse width. Similarly PSR J1740+1000 gives $r_f=0.12R_{LC}$.
Here we constrain $\zeta=80^{\circ}$ to $130^{\circ}$,
so that the corrected fit altitude (Equation \ref{eq:genCor} or Equation \ref{eq:zetaCor}) is 
$r^\prime_f=0.13R_{LC}$, again a small but statistically significant increase.

	For millisecond pulsars the effects can be larger. For example, \cite{k12} 
find that RVM fitting can be usefully applied to several 
recycled pulsars. PSR J1502-6752 (P=26.7\,ms) is a mildly recycled
pulsar for which the phase lag implies $r_f=0.2R_{LC}$. With no significant $\zeta$
constraints, we apply Equation (\ref{eq:genCor}) to infer a 16\% altitude increase
to $r^\prime_f=0.23R_{LC}$. Similarly PSR J1708-3506 ($P=4.5\,$ms) has a phase shift 
implying $0.19R_{LC}$, which we correct to $0.21-0.22 R_{LC}$. For this pulsar,
a naive application of the pulse width formula (\ref{eq:rW}) gives altitudes of 
$r_{W_{10}}\simeq0.65R_{LC}$ (10\% peak width). 
However, the increased $r^\prime_f$ and decreased pulse width height
from sweepback effects (Figure \ref{fig:totDirFitW}), along with additional current-induced
open zone growth, make it likely that the pulse width can be accommodated at the corrected height.

 \cite{k12} also report a RVM/BCW height $r_f= 0.44R_{LC}$ for
the P=2.7\,ms pulsar J1811-2404, along with well constrained viewing angles of
$\alpha=89.7^{\circ}$ and $\beta=21^{\circ}$. While our full analysis does
not cover this altitude, as Figure \ref{fig:SmapCorrected4} shows the corrections
of Equation (\ref{eq:zetaCor}) give a very high accuracy for orthogonal
rotators viewed near $90^\circ$.  Note in Figure \ref{fig:zetaResmagMin}, bottom
panel, that the correction function is nearly linear thus extrapolation to 
somewhat higher values may be justified.  Naively applying this correction we get
$r_f'=0.81R_{LC}$. We certainly cannot trust this value in detail since plasma 
effects and other perturbations may be relevant at such altitudes.  However, the
correction is certainly large and it brings the expected height up to an altitude 
where the very wide observed radio pulse, and the likely detection of emission 
from both open zones, can be easily accommodated.
Certainly simple RVM/BCW fitting is inadequate for this pulsar and one
should use a detailed model for the high altitude field geometry.

\bigskip
\bigskip

	Our exercise extends the range of utility of RVM-fit polarization
sweeps for inferring the altitudes of radio pulsar emission. For fit altitudes
less than $r_f = 0.25 R_{LC}$ the corrections are not large, but they are
systematic and, for high S/N data localizing the phase of maximum polarization
sweep, they can be highly significant. We thus believe it is worth applying our recommended
correction. For larger altitudes the corrections grow rapidly, but we caution
that as one approaches the light cylinder, current-induced distortions
should increase and, except for near-orthogonal rotators, one would expect the RVM
formulae to provide a poor fit in any case. Fitting to detailed numerical models
is then preferred. In all cases the dominant residual uncertainty is likely in
locating the phase of the radio pulse.
We also checked the use of absolute polarization axis position angles
and pulse width to constrain the emission height. Here the difficulties in
establishing the unperturbed $\psi_0$ and the expected distortions of the open zone
boundaries by currents, etc. make the estimates much less useful. Nevertheless,
we have shown that the effects of sweepback do go in the direction of reconciling
observed pulsar properties to a consistent emission height: larger heights are
inferred by a given $\Delta \psi$ shift and larger pulse widths can be accommodated
at a given height. We feel, however, that the corrections are less quantitative
than for $\Delta \phi$.

	In sum, since observers will continue to apply analytic RVM fits
to pulsar polarization data, by applying our recommended 
correction (Equation \ref{eq:zetaCor}), these results
can continue to give accurate height estimates to $\le 0.3 R_{LC}$. At higher heights
which will be common for millisecond pulsars, a fit to more detailed numerical
models is likely warranted.

\acknowledgements

This work was supported in part by NASA grants NNX10AP65G and NAS5-00147.

\end{document}